\documentclass[pdflatex,sn-mathphys-num]{sn-jnl}

\usepackage{graphicx}%
\usepackage{multirow}%
\usepackage{amsmath,amssymb,amsfonts}%
\usepackage{amsthm}%
\usepackage{physics}
\usepackage{mathrsfs}%
\usepackage[title]{appendix}%
\usepackage{xcolor}%
\usepackage{textcomp}%
\usepackage{manyfoot}%
\usepackage{booktabs}%
\usepackage{algorithm}%
\usepackage{algorithmicx}%
\usepackage{algpseudocode}%
\usepackage{listings}%
\theoremstyle{thmstyleone}%
\theoremstyle{thmstyletwo}%

\theoremstyle{thmstylethree}%

\begin{document}

\title[Cauchy-Schwarz and Bell Inequality Violations in Coupled Optomechanical Systems]{Cauchy-Schwarz and Bell Inequality Violations in Coupled Optomechanical Systems}

\author*[1]{\fnm{Joy} \sur{Ghosh}}\email{joyghos@kgpian.iitkgp.ac.in}

\author[2]{\fnm{Kapil} \sur{ Debnath}}

\author[1,3]{\fnm{Shailendra K.} \sur{Varshney}}

\affil*[1]{School of Nanoscience and Technology, Indian Institute of Technology Kharagpur, West Bengal, 721302, India}

\affil[2]{School of Natural and Computing Sciences, University of Aberdeen, Aberdeen AB24 3UE, UK}

\affil[3] {Department of Electronics and Electrical Communication Engineering, Indian Institute of Technology Kharagpur, West Bengal, 721302, India}


\abstract{Quantum destructive interference plays a subtle role in generating single photon Fock states that lead to fundamental classical inequalities violations. This work demonstrates two such violations, namely, (i) the Cauchy-Schwarz inequality and (ii) Bell nonlocality in a weakly perturbed coupled optomechanical system. Anticorrelation in antibunching occurs in this setup by coherently driving both the optical modes, resulting in second-order correlation functions less than the Poissonian limit. Our numerical simulations reveal that the antibunching effect can be tuned by adjusting the Kerr nonlinearity due to optomechanical coupling and photon hopping rates. The observed anticorrelation in photonic states can be attributed to interference between multiple transition pathways induced by the two distinct drives and corresponding phases. Additionally, we propose that violating the Cauchy-Schwarz inequality can serve as a precursor to stronger nonclassical tests, such as violations of the Bell inequality. These classical violations are significant for generating sub-Poissonian signals for quantum information processing tasks in hybrid systems based on single photon pairs.}


\maketitle

\section{Introduction}\label{sec1}

Since the 20th century, the scientific community has been bewildered about the quantum behavior of light, which is somewhat understood through pioneering works on optical coherence, where photon correlations have been identified as the fundamental resources of electromagnetic fields. Such photon correlations has made it possible to explain different nonclassical effects such as antibunching, squeezing, and entanglement \cite{agarwal2012quantum}. The most well-demonstrated examples include twin-beam generation \cite{bucker2011twin}, four-wave mixing \cite{marino2008violation}, parametric down-conversion \cite{luo2017chip}, resonance fluorescence \cite{PhysRevLett.39.691}, etc. The central argument for the proposal of nonclassical demonstration is the phenomenon, called anticorrelation in antibunching \cite{kheruntsyan2012violation}, which is easily measured by implementing a ubiquitous relation of mathematical physics and engineering called the Cauchy-Schwarz (CS) inequality \cite{steele2004cauchy}. The CS inequality provides a classical upper bound, which states that products of autocorrelations of two arbitrary fluctuating vectors are bounded by the squared expectation value of their cross-correlations, i.e., $|\langle \mathcal{AB} \rangle|\leq \sqrt{\langle \mathcal{A}^2 \rangle\langle \mathcal{B}^2 \rangle}$, where $\mathcal{A}$ and $\mathcal{B}$ are two random variables; any classical signals always obey this fundamental relation. However, the two-photon antibunching can violate this inequality, which is not accountable in classical optics \cite{reid1986violations}. Previously, the violation has been reported in a plethora of optical systems and atomic ensembles \cite{ansari1988violation,munoz2014violation,wu2009bright,trif2015photon,araujo2022observation}, as well as with matter-waves \cite{kheruntsyan2012violation}. The CS violation is also considered a major prediction of the spontaneous Hawking radiation in sonic black holes \cite{de2014violation}. The implication of CS violation is two-fold; firstly, it depicts the stronger quantum correlations between multimode bosonic systems, which are absent in the classical picture, and secondly, it implies the possibility of nonlocal effects encountered in the Clauser-Horne-Shimony-Holt (CHSH) framework \cite{kheruntsyan2012violation,reid1986violations}. The CHSH inequality \cite{clauser1969proposed} is a particular type of Bell inequality \cite{bell2004speakable} that falsified the idea of local realism advocated by  Einstein, Podolsky, and Rosen (EPR) in the hidden variable model \cite{einstein1935can}. Bell-CHSH violation has been reported in numerous platforms \cite{hensen2015loophole,palacios2010experimental,storz2023loophole,ansmann2009violation} which involve correlations between measurements on entangled particles that are usually represented by Bell states. Violations of the Bell-CHSH inequality have utmost importance in the current discourse of physical theories, providing evidence for the non-local phenomenon encountered in quantum mechanics. On a fundamental aspect, it is noteworthy to analyze CS and Bell-CHSH violations altogether in micro and nanomechanical systems for future quantum computation and communication protocols involving correlations at the single-photon-phonon level. 

Over the past decade, considerable efforts have been devoted to exploring the nonlinear interaction of nanoscale mechanical oscillators with optical cavities via the radiation pressure force of light, which gave birth to the field of optomechanics \cite{aspelmeyer2014cavity}. The ground state cooling \cite{guo2019feedback}, normal mode splitting \cite{dobrindt2008parametric}, entanglement between mirror and light \cite{vitali2007optomechanical}, and squeezing of mechanical oscillators \cite{PhysRevA.109.023512} showed breakthroughs, which brings optomechanical (OM) systems a considerable participant while investigating the nonlinear quantum regime on a mesoscopic scale \cite{barzanjeh2022optomechanics}.
The antibunching effect, typically associated with photons, has now been extended to phononic modes as well \cite{humphries2023phonon}. It is widely recognized that higher-order nonlinearity is crucial for generating antibunching \cite{PhysRevLett.104.183601}. To achieve this, various mechanisms for photon and phonon blockades have been proposed, including two-level systems \cite{wang2015tunable}, parametric amplification \cite{wang2020enhanced}, spinning resonators \cite{huang2018nonreciprocal}, $\mathcal{PT}$-symmetric effects \cite{wang2019distinguishing}, quadratic coupling \cite{liao2013photon,xie2017phonon}, qubit-based unconventional phonon blockade \cite{xu2016phonon} and magnon blockade \cite{fan2023nonclassical}, etc. However, the real challenge in photon blockade parameter estimation is to examine whether the effects of weak nonlinearity or strong coupling coefficient are more dominant in a given setup. 

This paper presents the CS and Bell-CHSH inequalities violation based on the strong photon antibunching in a coupled opto-mechanical system. The second-order correlation functions are obtained by both numerical simulations of the Lindblad master equation and analytical solutions of the Schrödinger equation. We find a good agreement between both approaches, resulting in photon blockade in the coupled OM system while driven by two weak control pumps. We also examine the intracavity statistical properties of photons using cross-correlation functions by regulating the amplitudes and relative phases of the external coherent fields. Consequently, a single-photon Fock state is predicted via multiple pathways of destructive interference \cite{bamba2011origin} in weakly nonlinear and strongly coupled parametric conditions. The hypothesis of EPR-like correlations, which may manifest upon CS inequality violation (as discussed in \cite{kheruntsyan2012violation}), is reflected in our findings under resonant excitation of both cavities. Additionally, the sub-Poissonian statistics and squeezing of the coherent fields have been investigated, which are highly relevant for quantum optical applications.

The work is organized as follows: in Sec. \ref{sec2}, the effective Hamiltonian of the system is presented. The numerical modelling of the correlation functions with relation to the Cauchy-Schwarz inequality is provided in Sec. \ref{sec3}. The simulation results are presented in Sec. \ref{sec4}, and an analytical description of this model is provided in Sec. \ref{sec5}. Lastly, a brief discussion is provided about the fidelity of Bell-type states associated with the classical violation in Sec. \ref{sec6}, which concludes the work.

\section{Theoretical model}\label{sec2}

\begin{figure}[htbp]
\includegraphics[width=\linewidth]{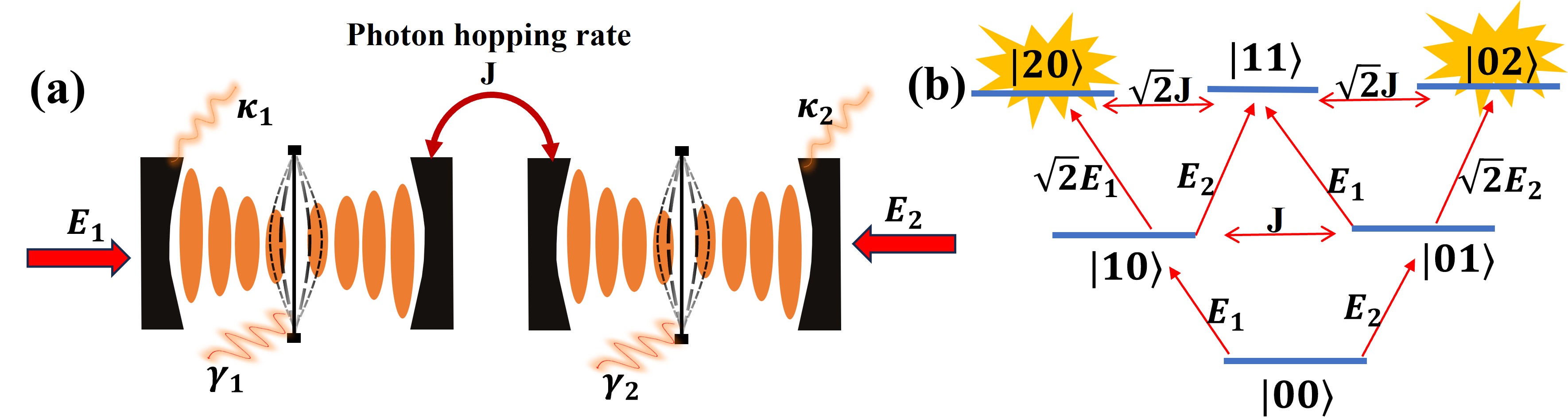}
\caption{(a) Schematic of two coupled optomechanical systems consisting of a thin membrane in the middle. The cavity is excited by a control pump with amplitude $E_{j}$ with phases $\theta_{j}$. The corresponding damping rates of the cavity and membrane are $\kappa_j$ and $\gamma_j$. $J$ denotes the optical coupling between the two systems. (b) Schematic representation of transition pathways of multiple Fock states yielding single photon blockade due to destructive quantum interference.}
\label{fig_1}
\end{figure}

We consider a coupled Fabry–Pérot optomechanical system, consisting of two laser-driven cavities that cause coherent vibrations in the mechanical oscillator situated in one of the nodes (or anti-nodes) of the cavity, as shown in Fig. \ref{fig_1}(a). The cavity frequencies $\omega_{1,2}$ are assumed to be linearly dependent on the displacement of the oscillator due to the optical radiation pressure force exerted by the laser amplitude $E_{1,2}$ (of frequency $\omega_{L}$). The Hamiltonian of the system can be expressed as (taking \(\hbar = 1\))
\begin{eqnarray}
\mathcal{H}&=&\omega_{1} a_1^\dagger a_1 + \omega_{2} a_2^\dagger a_2 + \omega_{m} b_1^\dagger b_1 + \omega_{m} b_2^\dagger b_2 + J (a_1^\dagger a_2 + a_1 a_2^\dagger)- g a_1^\dagger a_1 (b_1^\dagger + b_1) \nonumber\\&&- g a_2^\dagger a_2 (b_2^\dagger + b_2)+ E_1 a_1^\dagger e^{-i\omega_{L} t} + E_2 a_2^\dagger e^{-i\omega_{L} t} +\text{hc},
\label{eq1}
\end{eqnarray}
where $\hat{a}^{\dagger}(\hat{a})$ and $\hat{b}^{\dagger}(\hat{b})$ are the creation (annihilation) operators associated with the optical and mechanical modes, respectively. For simplicity, identical mechanical oscillators' frequency $\omega_{m}$ and single-photon-phonon coupling strength $g$ are considered for both OM systems. Furthermore, $J$ denotes the evanescently coupled optical modes' photon hopping rate for the two OM systems. The laser amplitude $E_j$ ($j=1,2$) is related to the input optical power $\mathcal{P}_{in}$ and cavity decay rate $\kappa_j$ as $|E_j|=\sqrt{\mathcal{P}_{in}\kappa_j/\omega_L}$. By applying the unitary transformation $V = \exp\left[-i \omega_{L} t (a_1^\dagger a_1 - a_2^\dagger a_2)\right]$ with respect to rotating laser frequency, the Hamiltonian becomes time independent as
\begin{eqnarray}
\mathcal{H'} &=& \Delta_1 a_1^\dagger a_1 + \Delta_2 a_2^\dagger a_2 + \omega_m b_1^\dagger b_1 + \omega_m b_2^\dagger b_2 + J (a_1^\dagger a_2 + a_1 a_2^\dagger)- g a_1^\dagger a_1 (b_1^\dagger + b_1) \nonumber\\&&- g a_2^\dagger a_2 (b_2^\dagger + b_2)+ E_1 a_1^\dagger + E_2 a_2^\dagger +\text{hc},
\label{eq2}
\end{eqnarray}
The detuning between cavity frequency and driving laser frequency is $\Delta_j=\omega_j-\omega_L$. This Hamiltonian can be further transformed into the mechanical oscillators' displacement picture, as the present work focuses on photon blockade effects in a weakly perturbed optomechanical system. The unitary transformation $V' =   \exp \sum_{j=1,2} \frac{g}{\omega_m}a_j^\dagger a_j (b_j^\dagger - b_j)$ diagonalizes $\mathcal{H''}\rightarrow V'^\dagger \mathcal{H'}V'$ as
\begin{eqnarray}
\mathcal{H''} &=&  \mathcal{H}_m+\sum_{j=1,2} \left[ \Delta_j a_j^\dagger a_j - U a_j^\dagger a_j a_j^\dagger a_j + (E_j e^{i\theta_j} a_j^\dagger
+ E_j e^{-i\theta_j} a_j) \right]  
 \nonumber\\&&+ J(a_1^\dagger a_2 + a_2^\dagger a_1),
\label{eq3}
\end{eqnarray}
where $\mathcal{H}_m=\sum_{j=1,2} \omega_m b_j^\dagger b_j$ represents the mechanical resonator's motion decoupled from the original system. As a result, an effective Kerr-like nonlinear interaction \( U = \frac{g^2}{\omega_m} \) arises due to OM coupling. But the effective Kerr nonlinear model is only valid for the weak OM coupling condition $\omega_m > g$; all higher order corrections are neglected under the Taylor series approximation of the exponential. This reduced Hamiltonian of Eq. (\ref{eq3}) only depicts the optical signatures of the system, which is similar to coupled photonic molecules \cite{bamba2011origin}. Contrary to the photonic molecules, here, the Kerr coefficient is controlled only by the properties of the mechanical oscillator. Additionally, Fig. \ref{fig_1}(b) depicts the Fock states in a truncated Hilbert space of this coupled OM system, which is essential for the photon blockade mechanism discussed subsequently.

\section{Lindblad master equation, second-order correlation functions and Cauchy-Schwarz inequality}\label{sec3}

The dynamical behavior of the effective Hamiltonian of Eq. (\ref{eq3}) can be obtained by numerically simulating the Lindblad master equation given by
\begin{equation}
    \partial_t \rho = -i[\mathcal{H''},\rho]+\sum^2_j\kappa_j\mathcal{L}(\hat{a}_j)\rho+(n_{\text{th}}+1)\gamma_j\mathcal{L}(\hat{b}_j)\rho+n_{\text{th}}\gamma_j\mathcal{L}(\hat{b}_j^\dagger)\rho,
\label{eq4}
\end{equation}
where $\rho$ is the density matrix and $n_{\text{th}}=\frac{a}{\exp(\frac{\hbar\omega_{m}}{k_B T})-1}$ is the mean thermal excitation number at bath temperature $T$ with $k_B$ the Boltzmann constant. Also, $\mathcal{L}(\hat{a}_j)=\hat{a}_j\rho\hat{a}_j^\dagger-\frac{1}{2}\{\hat{a}_j^\dagger\hat{a}_j,\rho\}$ is known as the Lindbladian operator and $\{\cdot\}$ representing anticommutation operation. Eq. (\ref{eq4}) contains the necessary information on the dissipation of the optical and mechanical resonators' motion, which is dependent on the reservoir temperature $T$, depicting decoherence. By solving the master equation, the photon statistics can be obtained by the following time-delayed second-order correlation functions
\begin{eqnarray}
    g_{a_j}^2(\tau) &=& \frac{\text{Tr}[\hat{a}_j^\dagger\hat{a}_j^\dagger\hat{a}_j\hat{a}_j\rho]}{\text{Tr}[\hat{a}_j^\dagger\hat{a}_j\rho]^2},
    \label{eq5}\\
    g_{a_ia_j}^2(\tau) &=& \frac{\text{Tr}[\hat{a}_i^\dagger\hat{a}_j^\dagger\hat{a}_j\hat{a}_i\rho]}{\text{Tr}[\hat{a}_i^\dagger\hat{a}_i\rho]\text{Tr}[\hat{a}_j^\dagger\hat{a}_j\rho]},
    \label{eq6}
\end{eqnarray}
where Tr$[\cdot]$ denotes the trace operation and $\hat{n}_j=\hat{a}_j^\dagger\hat{a}_j$ gives the intracavity photon number of the jth cavity. The equal time correlation functions in Eq.(\ref{eq5}) and Eq.(\ref{eq6}) are frequently used to characterize the joint probability of detecting two single photons, which measures the nonclassical effects often described as sub-Poissonian. However, for the classical case, the correlation functions should always obey the following inequalities, (i) $g_{a_j}^2(0)\geqslant 1$, (ii) $g_{a_j}^2(\tau)\leqslant g_{a_j}^2(0)$, and (iii) $g_{a_ia_j}^2(0)\leqslant\sqrt{g_{a_i}^2(0)g_{a_j}^2(0)}$, where $i\neq j$ \cite{miranowicz2010testing}. 

The first condition describes the coherent (thermal) nature of classical fields, and its violation indicates the sub-Poissonian statistics associated with photon antibunching. The limit $g_{a_j}^2(0)\rightarrow0$ corresponds to the ideal blockade phenomena in which only single photon Fock states are excited. The violation of the second and third inequalities displays the effect of anticorrelation in antibunching \cite{kheruntsyan2012violation}, which results in the violation of the Cauchy-Schwarz inequality. The cross-correlation function $ g_{a_ia_j}^2(0)<1$ indicates the competition of single photon excitation in both cavities, which is unlikely to happen simultaneously. In contrast, $ g_{ij}^2(0)>1$ gives the perfect photon pair generation in the two coupled OM systems. It is important to note that the constraint on the cross-correlation function imposed by the third inequality is much stronger, usually known as the classical CSI. To demonstrate its violation, the intracavity photon covariance is calculated as $\mathcal{C}=\langle\hat{n}_1\hat{n}_2 \rangle -\langle\hat{n}_1\rangle\langle\hat{n}_2\rangle$. If there exists a finite correlation between the generated photon pair, i.e., $\mathcal{C}\neq0$, the classical CSI obeys 
\begin{eqnarray}
    \mathcal{C}=\frac{g_{a_ia_j}^2(0)}{\sqrt {g_{a_i}^2(0)g_{a_j}^2(0)}}\leqslant 1,
    \label{eq7}
\end{eqnarray}
Otherwise, quantum correlations exist if the inequality is violated. This violation is experimentally measurable by obtaining the output spectrum of the system using the widely popular Hanbury, Brown, and Twiss type experimental setup \cite{brown1956correlation,cohen2015phonon}. The sub-Poissonian statistics characterized by the second-order correlation function can also be linked to the Fano factor, from which the optomechanical phonon laser is achievable \cite{lorch2015sub}. Additionally, in the case of Gaussian states, the classically forbidden values of the correlation functions are mainly observed as a consequence of optimized amplitude squeezing \cite{lemonde2014antibunching}.

\begin{figure}[htbp]
\includegraphics[width=\linewidth]{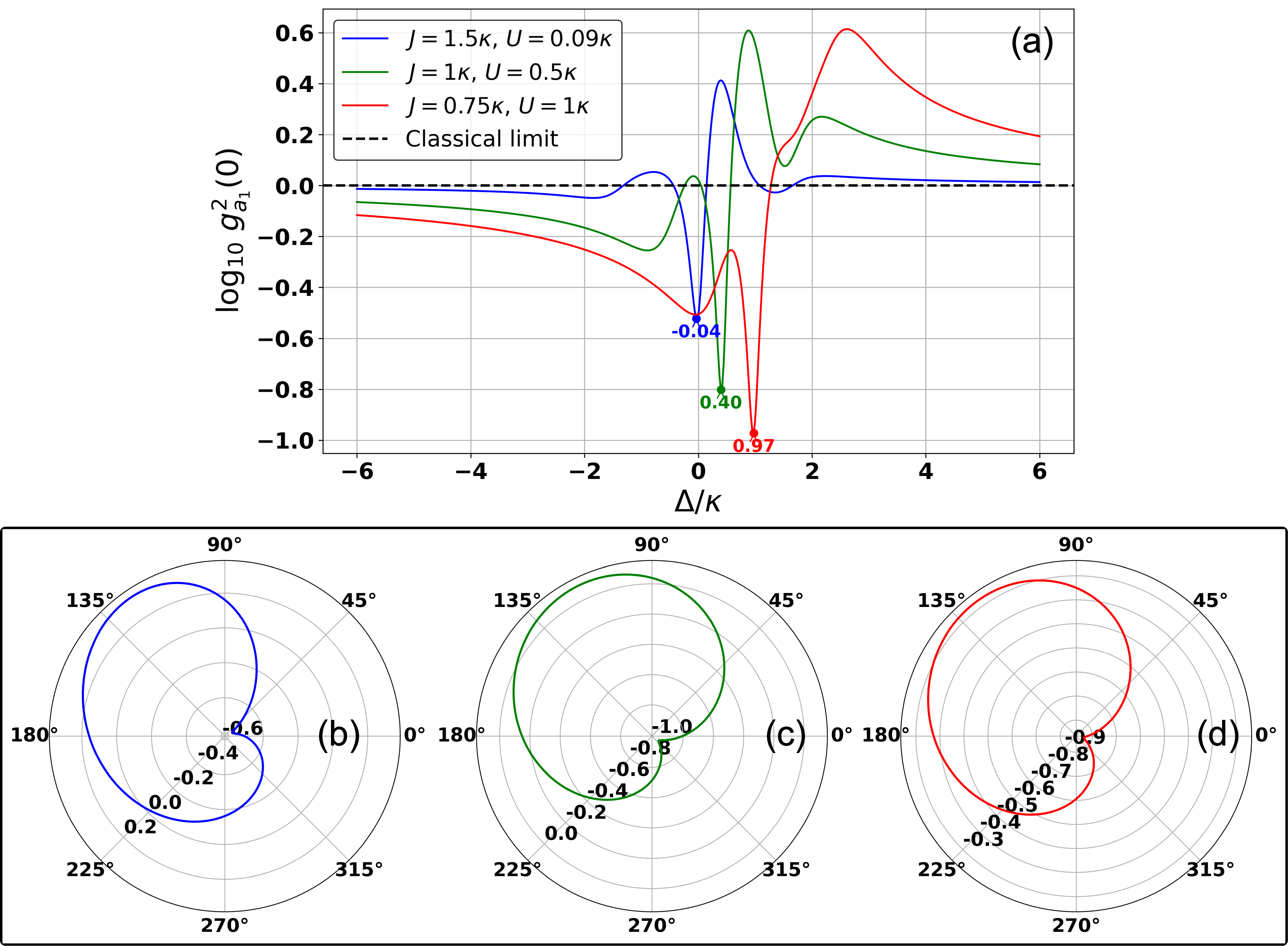}
\caption{(a) The equal-time second-order correlation function $g^{(2)}_{a_1}(0)$ (Log scale) as a function of normalized detuning $\Delta/\kappa$ for three different parameter regimes: $J = 1.5\kappa, U = 0.09\kappa$ (blue), $J = \kappa, U = 0.5\kappa$ (green), and $J = 0.75\kappa, U = \kappa$ (red). The black dashed line marks the classical threshold $g_{a_1}^{(2)} = 1$. Panels (b)--(d) display the corresponding photon blockade with the phase difference of the coherent fields $\theta$ in polar coordinates for each case.}
\label{fig_2}
\end{figure}
\begin{figure}[htbp]
\includegraphics[width=\linewidth]{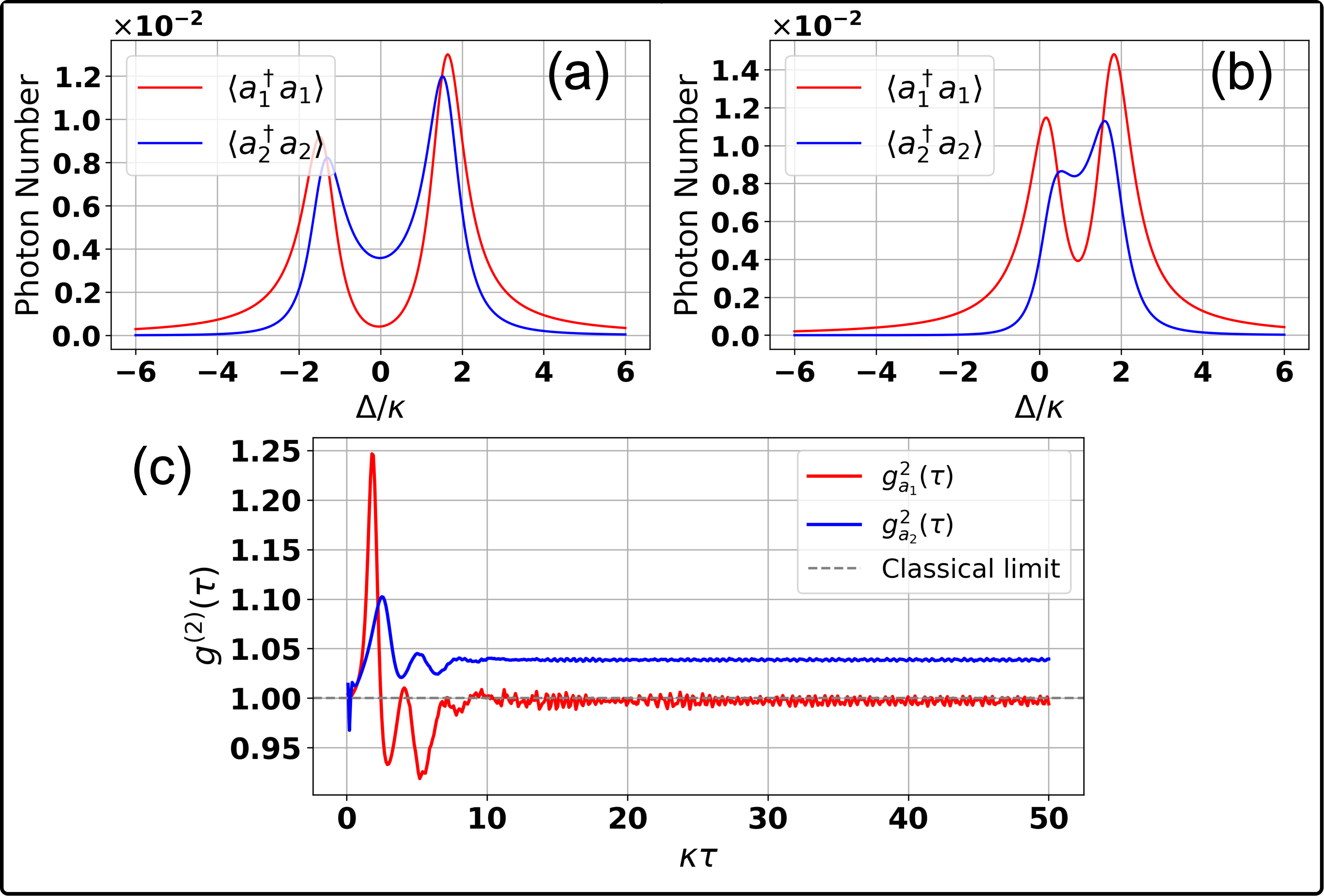}
\caption{The steady-state photon numbers as functions of normalized detuning \( \Delta/\kappa \), for parameters (a) \(J = 1.5\kappa, U = 0.09\kappa \) and (b) \(J = 0.75\kappa,  U = \kappa \), respectively. Both cases exhibit two distinct resonance peaks due to hybridization via the inter-cavity coupling \( J \). (c) Time evolution of the second-order correlation functions, revealing the violation of the second classical inequality.}
\label{fig_3}
\end{figure}

\section{Numerical results of single photon blockade}\label{sec4}

The effective Hamiltonian of Eq. (\ref{eq3}) exhibits symmetry in the Fock space; thus, we only show the numerical simulation of the second-order correlation function for $g^2_{a_1}(0)$ in Fig. \ref{fig_2}. The photon statistics are obtained by solving the Lindblad master equation from the steady-state density matrix of Eq. (\ref{eq4}), where the numerical solver mesolve in QuTip \cite{johansson2012qutip} is used. In simulations, we assume the identical parametric conditions, $U_1=U_2=U$, $\kappa_1=\kappa_2=\kappa$, $\gamma_1=\gamma_2=\gamma$. However, the field strengths are taken as $ E_1=0.1\kappa, E_2 = 0.01\kappa$, which can be controllable and vice versa as well. This parameter approximation helps in solving the system dynamics analytically, where the thermal phonon occupancy is considered zero in all numerical calculations. The quality factor of the mechanical mode is $\omega_m/\gamma_{} = 10^{6}$ and the frequency is fixed at $\omega_m = 100\kappa$ MHz such that OM coupling can be tuned between $g =3\kappa\sim 10\kappa$ MHz to control the effects of the Kerr coefficient $U$. The parameters of this model are taken according to the good cavity limit mentioned in the foundational works \cite{rabl2011photon,nunnenkamp2011single}, operated in the resolved sideband regime. 

The second-order correlation function $g^2_{a_1}(0)$ has been calculated from Eq. (\ref{eq5}) considering steady state case and the resonant condition $\Delta_1=\Delta_2=\Delta$. The photon hopping rate $J$ is tailored to achieve weak to strong coupling regimes in order to obtain Hong-Ou-Mandel dips, as shown in Fig. \ref{fig_2}(a). The antibunching phenomenon is characterized by the dips of the second-order photon correlation function observed at $\Delta = -0.04\kappa, 0.4\kappa, 0.97\kappa$ in Fig. \ref{fig_2}(a), indicating violation of the first inequality ($\log g_{a_1}^2(0)<0$). The solid red curve shows the most significant suppression of $g^{2}_{a_1}(0)\approx 0.1 $ at $J=0.75\kappa$ and $U=\kappa$. This strong antibunching (characterized by a minimum $g^{2}_{a_1}(0)$) is a hallmark of pronounced nonclassical behavior, originating from the interplay of strong Kerr nonlinearity and moderate coupling.

In addition to the amplitudes, the relative phase difference of the two coherent fields plays a significant role in obtaining photon blockade. The lower panel of Fig. \ref{fig_2}(b)--(d) represents the corresponding antibunching characteristics for the variation in $\theta =\theta_1-\theta_2$. These polar representations reveal that $g^{2}_{a_1}(0)$ exhibits a pronounced dependence on the relative drive phase. Notably, the strongest antibunching occurs at optimal phase-matching conditions in each scenario. This phase-sensitive modulation of photon statistics stems from multiple pathway destructive interference (shown in Fig. \ref{fig_1}(b)) effects between the coherently driven modes.

Next, we demonstrate the steady-state mean photon numbers $\langle a_1^\dagger a_1 \rangle$ and $\langle a_2^\dagger a_2 \rangle$ for both systems as a function of the normalized detuning $\Delta/\kappa$, when Kerr nonlinearities are $U = 0.09\kappa$ (Fig. \ref{fig_3}(a)) and $U = \kappa$ (Fig. \ref{fig_3}(b)), respectively. In both cases, two distinct resonance peaks are observed, corresponding to normal-mode splitting due to coherent tunneling between the two optical cavities via the coupling strength $J$. The photon number is significantly smaller at zero detuning, which is evident in Fig. \ref{fig_3}(a) for the $a_1$ mode. This indicates a regime where two excitation pathways, direct and indirect via inter-cavity coupling, interfere destructively. The Fano-type antiresonance arises from the combined influence of Kerr nonlinearity and an asymmetric drive, with $E_1\gg E_2$. The difference is more visible in Fig. \ref{fig_3}(b), indicating that higher nonlinearity suppresses energy transfer to cavity 2 more strongly. This phenomenon is also known as the number-squeezing effect, which is due to the intracavity statistical mixture of photons \cite{reid1986violations}.

Furthermore, the temporal evolution of the second-order correlation function is measured by the joint probability of detecting one photon at time $t=0$ and the next emitted photon at time $t=\tau$ by solving Eq. (\ref{eq5}) and Eq. (\ref{eq6}). The dynamics of $\log_{10}g_{a_i}^2(\tau)$ are illustrated in Fig. \ref{fig_3} for a single case $J=0.75\kappa$ and $U=\kappa$ only, to validate the violation of the second inequality. The temporal evolution of $g_{a_i}^2(\tau)>g_{a_i}^2(0)$ exhibits a fluctuation for finite time delay $\tau$ in the transient stage, which stabilizes to steady values around or slightly above the classical limit due to damping and decoherence. The external field strengths are fixed, possessing a phase difference $\theta=0$ and the detuning $\Delta=0.97\kappa$. Fig. \ref{fig_3}(c) reflects that the low-power parametric process generates significant quantum correlations in cavity optomechanics \cite{huang2010normal}.

\section{Analytical methods}\label{sec5}
In this section, we present the analytical findings of the classical inequality violations and compare them with the numerical simulations of the second-order correlation function demonstrated previously. The CSI violations, along with the Bell-CHSH inequality violation, are discussed in detail, and the equivalence of classical violation is illustrated. 

\subsection{Steady-state solution}

The analytical expressions of the second-order correlation function $g^2_{a_i}(\tau)$ are not directly solvable from the master equation Eq. (\ref{eq4}). To better understand the physical mechanism of photon blockade, an analytical method is presented to breach the classical inequalities and antibunching phenomena. The steady-state second-order correlation function can be calculated via Schrödinger's equation in the truncated Fock space with a weak field approximation, i.e, $E_{1,2}<\kappa$, where the possibility of multi-particle excitation can be controlled by optimizing the field's amplitude and relative phases. If the membrane is cooled to the ground state initially $(n_{\text{th}}\approx0)$, the weak coherent fields can be treated as a perturbation. The wave function can be expressed as \cite{bamba2011origin}
\begin{eqnarray}
\ket{\psi}=\sum^{n_1+n_2\leq2}_{n_1,n_2}C_{n_1,n_2}\ket{n_1,n_2},
    \label{eq8}
\end{eqnarray}
where $C_{n_1,n_2}$ is the amplitude of the Fock state $\ket{n_1}\bigotimes\ket{n_2}$ with occupation probability $|C_{n_1,n_2}|^2$, satisfying, $|C_{00}|\gg |C_{10}|$,$|C_{01}|\gg |C_{11}|$,$|C_{20}|$,$|C_{02}|$. Considering dissipation factors of the two OM cavities into account, the effective Hamiltonian has the following non-Hermitian form written as, $\mathcal{H}_{eff}=\mathcal{H''}-i\kappa_1/2\hat{a}_1^\dagger\hat{a}_1-i\kappa_2/2\hat{a}_2^\dagger\hat{a}_2$. Based on the Schrödinger's equation i.e. $i \frac{d}{dt}\ket{\psi}=\mathcal{H}_{eff}\ket{\psi}$, the dynamical evolution of the transition probabilities are found as
\begin{eqnarray}
i \frac{d}{dt} c_{00} &=& E_1e^{i\theta_1} c_{10} + E_2e^{i\theta_2} c_{01},
\label{eq9}\\
i \frac{d}{dt} c_{10} &=& (\Delta_1 - i \frac{\kappa_1}{2} - U_1) c_{10} + J c_{01} + \sqrt{2} E_1e^{i\theta_1} c_{20} + E_1e^{i\theta_1} c_{00} + E_2e^{i\theta_2} c_{11},
\label{eq10}\\
i \frac{d}{dt} c_{01} &=& (\Delta_2 - i \frac{\kappa_2}{2} - U_2) c_{01} + J c_{10} + \sqrt{2} E_2e^{i\theta_2} c_{20} + E_2e^{i\theta_2} c_{00} + E_1e^{i\theta_1} c_{11},
\label{eq11}\\
i \frac{d}{dt} c_{11} &=& (\Delta_1 + \Delta_2 - i \frac{\kappa_1}{2} -i\frac{\kappa_2}{2} - U_1 -U_2) c_{11} + \sqrt{2}J (c_{20}+c_{02}) + E_1e^{i\theta_1} c_{01} \nonumber\\&& + E_2e^{i\theta_2} c_{10},
    \label{eq12}\\
i \frac{d}{dt} c_{20} &=& 2(\Delta_1 - i \frac{\kappa_1}{2} - 2U_1) c_{20} + \sqrt{2}J c_{11} + \sqrt{2} E_1e^{i\theta_1} c_{10},
   \label{eq13} \\
i \frac{d}{dt} c_{02} &=& 2(\Delta_2 - i \frac{\kappa_2}{2} - 2U_2) c_{02} + \sqrt{2}J c_{11} + \sqrt{2} E_2e^{i\theta_2} c_{01}.
   \label{eq14}
\end{eqnarray}
It is found that Eq. (\ref{eq9}) dictating probability amplitude of the ground state is always approximately satisfied, suggesting $|C_{00}|^2\approx 1$. From the different transitions among the Fock states, the quantum interference effect can be readily seen (depicted in Fig. \ref{fig_1}(b)). The different routes exist for two-photon transition are i.e. the direct path (i) $\ket{00}\xrightarrow[]{E_1}\ket{10}\xrightarrow{\sqrt{2}E_1}\ket{20}$, and the tunneling mediated path (ii) $\ket{00}\xrightarrow{E_1}\ket{10}\xrightarrow{J}\ket{01}(\xrightarrow{\sqrt{2}E_2}\ket{02}\xrightarrow{\sqrt{2}J}\ket{11})\xrightarrow{E_1}\ket{11}\xrightarrow{\sqrt{2}J}\ket{20}$, where quantum destructive interference is achieved by gaining additional phases from different paths. This blocks the two-photon excitation, resulting in the probability of detecting $\ket{20},\ket{02}$ becoming zero.  By doing steady-state analysis, the Eq. (\ref{eq10}) to Eq. (\ref{eq14}) can be solved iteratively. The one particle states are approximately found as
\begin{eqnarray}
    (\Delta_1'-U_1) C_{10} + J C_{01} &=& -E_1e^{i\theta_1},
    \label{eq15}\\
    (\Delta_2'-U_2) C_{01} + J C_{10} &=& -E_2e^{i\theta_2},
    \label{eq16}
\end{eqnarray}
where $\Delta'_j = \Delta_j-i\kappa/2$ and $C_{10}$, $C_{01}$ are obtained as
\begin{eqnarray}
    C_{10} &=& \frac{J E_2e^{i\theta_2}-E_1 e^{i\theta_1} (\Delta_2'-U_2)}{(\Delta_1'-U_1)( \Delta_2' -U_2)- J^2},
    \label{eq17}\\
    C_{01} &=& \frac{JE_1e^{i\theta_1} - E_2e^{i\theta_2}(\Delta_1'-U_1)}{(\Delta_1'-U_1)( \Delta_2' -U_2)- J^2}.
    \label{eq18}
\end{eqnarray}
The coefficients for two-particle states follows
\begin{eqnarray}
    2(\Delta_1' - 2U_1) C_{20} + \sqrt{2} J C_{11} &=& -\sqrt{2}E_1e^{i\theta_1} C_{10},
    \label{eq19}\\
    2(\Delta_2' - 2U_2) C_{02} + \sqrt{2} J C_{11} &=& -\sqrt{2}E_2e^{i\theta_2} C_{01},
    \label{eq20}\\
    (\Delta_1' + \Delta_2' - U_1 - U_2) C_{11} + \sqrt{2} J C_{20} + \sqrt{2} J  C_{02} &=& -E_1e^{i\theta_1} C_{01} - E_2 e^{i\theta_2} C_{10}.
    \label{eq21}
\end{eqnarray}
The auto-correlation and cross-correlation functions are computed as
\begin{eqnarray}
    g^2_{a_1}(0)&=& \frac{\langle\hat{a}_1^\dagger\hat{a}_1^\dagger\hat{a}_1\hat{a}_1\rangle}{\langle\hat{a}_1^\dagger\hat{a}_1\rangle^2} =\frac{2|C_{20}|^2}{(|C_{10}|^2+|C_{11}|^2+2|C_{20}|^2)^2} \approx \frac{2|C_{20}|^2}{|C_{10}|^4},
    \label{eq22}\\
    g^2_{a_2}(0)&=&\frac{\langle\hat{a}_2^\dagger\hat{a}_2^\dagger\hat{a}_2\hat{a}_2\rangle}{\langle\hat{a}_2^\dagger\hat{a}_2\rangle^2}=\frac{2|C_{02}|^2}{(|C_{01}|^2+|C_{11}|^2+2|C_{02}|^2)^2}\approx \frac{2|C_{02}|^2}{|C_{01}|^4},
    \label{eq23}\\
    g^2_{a_1a_2}(0)&=&\frac{\langle\hat{a}_1^\dagger\hat{a}_2^\dagger\hat{a}_2\hat{a}_1\rangle}{\langle\hat{a}_1^\dagger\hat{a}_1\rangle\langle\hat{a}_2^\dagger\hat{a}_2\rangle}=\frac{|C_{11}|^2}{(|C_{10}|^2+|C_{11}|^2+2|C_{20}|^2)
    \times(|C_{01}|^2+|C_{11}|^2+2|C_{02}|^2)}\nonumber\\&&\approx \frac{|C_{11}|^2}{|C_{10}|\times|C_{01}|}.
    \label{eq24}
\end{eqnarray}
The nonclassical result ($g^2_{a_j}(0)<1$) can be obtained by making  $C_{20},C_{02}\approx0$ in the probability transition equations. The optimal conditions for the photon blockade are the nontrivial solutions obtained from the determinant of the coefficient matrix
\begin{eqnarray}
   \det 
\begin{bmatrix}
2(\Delta_1' - 2U_1) & 0 & \sqrt{2} J \\
0 & 2(\Delta_2' - 2U_2) & \sqrt{2} J \\
\sqrt{2} J & \sqrt{2} J & (\Delta_1' + \Delta_2' - U_1 - U_2)
\end{bmatrix}
    = 0.
\end{eqnarray}
If the higher-order terms in the equation governing \( c_{20} \), are neglected, $ 2\left(\Delta_1 - i \frac{\kappa_1}{2} - 2U_1\right) c_{20} + \sqrt{2}J c_{11} + \sqrt{2} E_1 e^{i\theta_1} c_{10} = 0$ leads to $c_{20} = \frac{-J c_{11} - E_1 e^{i\theta_1} c_{10}}{\sqrt{2}(\Delta_1 - i \frac{\kappa_1}{2} - 2U_1)}$. To enforce photon blockade, we set \( c_{20} = 0 \), which yields $c_{11} = -\frac{E_1}{J} e^{i\theta_1} c_{10}$.
After substituting into the equation for \( c_{11} \), and simplifying (neglecting \( c_{02} \)) leads to the consistency condition:
$0 = -\chi \cdot \frac{E_1}{J} e^{i\theta_1} c_{10} + E_1 e^{i\theta_1} c_{01} + E_2 e^{i\theta_2} c_{10}$, where $
\chi = \Delta_1 + \Delta_2 - U_1 - U_2 - i \frac{\kappa_1 + \kappa_2}{2}$. By algebraic elimination and substitution, we arrive at the optimal blockade condition $\left( \Delta_1 - 2U_1 - i \frac{\kappa_1}{2} \right)
\left( \Delta_1 + \Delta_2 - U_1 - U_2 - i \frac{\kappa_1 + \kappa_2}{2} \right)
= \frac{E_1^2}{J^2} \left( \Delta_2 - U_2 - i \frac{\kappa_2}{2} \right)$.
 Here, we choose the conventional single photon resonance $\Delta=U$, while the far-off resonant term $\Delta=2U$ is neglected. The coupling-induced interference caused by $J$ can be tuned so that the numerator is suppressed (interference) and the denominator is enhanced (resonance for single-photon transition, detuned for two-photon) in Eq. (\ref{eq22})--(\ref{eq24}). 

\begin{figure}[htbp]
\includegraphics[width=\linewidth]{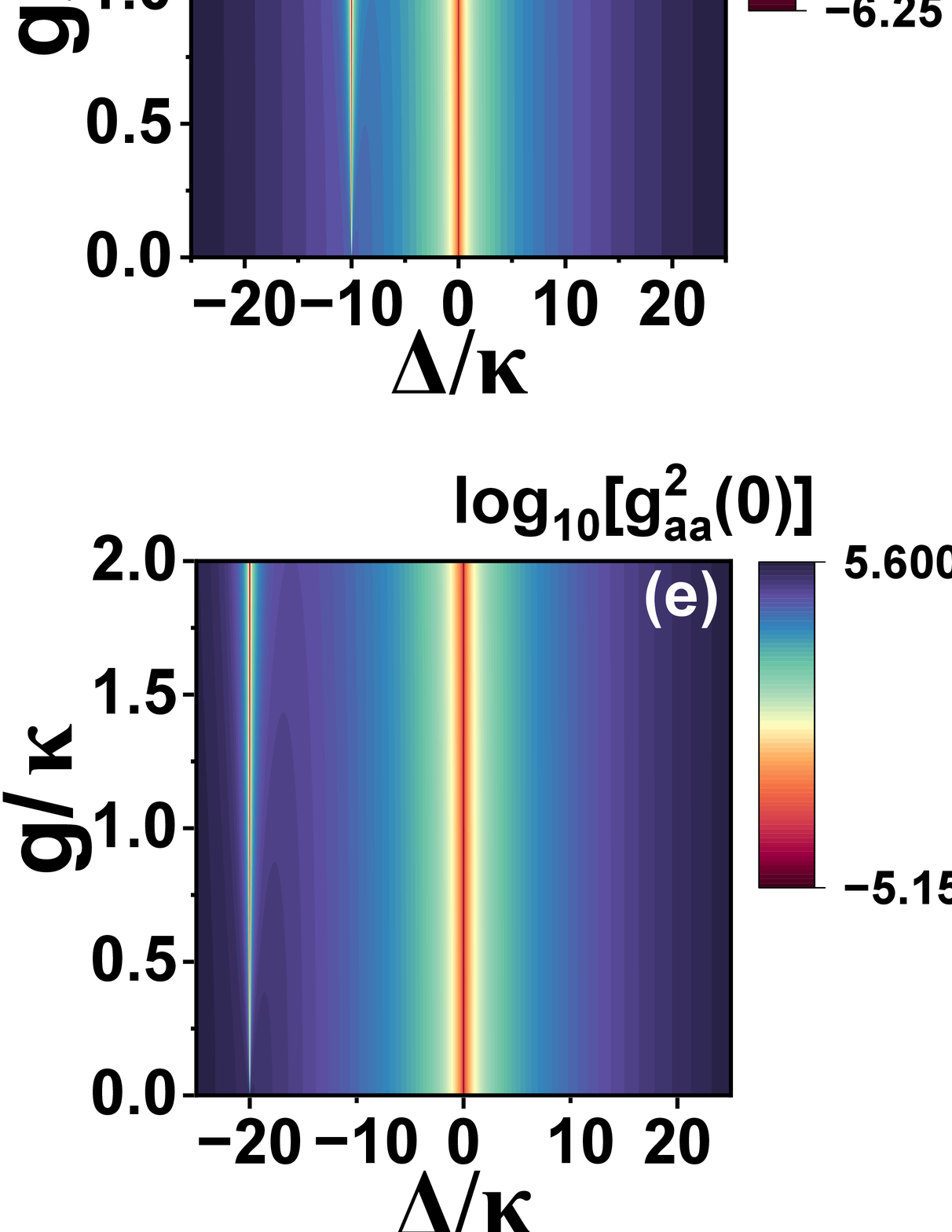}
\caption{Analytical results of the second-order correlation functions (logarithmic scale) $g^{(2)}_{a_1}(0)$ (red), $g^{(2)}_{a_2}(0)$ (blue), and the non-classicality witness CS violation $\mathcal{C}$ (yellow) as a function of normalized detuning $\Delta/\kappa$, for three different regimes of Kerr nonlinearity $U$ and inter-cavity coupling $J$: (a) $U = 0.09\kappa$, $J = 1.5\kappa$; (c) $U = 0.5\kappa$, $J = \kappa$; and (e) $U = \kappa$, $J = 0.75\kappa$. Corresponding contour plots in (b), (d), and (f) show the dependence of $\log_{10}g^{(2)}_{a_1}(0)$ on the detuning $\Delta/\kappa$ and Kerr nonlinearity $U/\kappa$, illustrating regions of strong photon antibunching while keeping other parameters the same.}
\label{fig_4}
\end{figure}

\subsection{Cauchy-Schwarz and Bell-CHSH inequality violations}

In the analytical derivations mentioned in the previous section, mathematical expressions yielding second-order correlation functions were obtained for both OM systems. The Cauchy-Schwarz violation, characterized as a nonclassical witness defined in Eq. (\ref{eq7}), can be expressed in terms of transition probability coefficients as
\begin{eqnarray}
    \mathcal{C}=\frac{|C_{11}|^2}{2|C_{20}|\times|C_{02}|},
    \label{eq26}
\end{eqnarray}
Similarly, the Bell-CHSH parameter can be derived as 
\begin{eqnarray}
    \mathcal{B}=\frac{1}{\sqrt{2}}\abs{\frac{2\abs{C_{20}}^2+2\abs{C_{02}}^2-4\abs{C_{20}}\abs{C_{20}}-4\abs{C_{11}}^2}{\abs{C_{20}}^2+\abs{C_{02}}^2+\abs{C_{11}}^2}}.
    \label{eq27}
\end{eqnarray}

The detailed calculation of the Bell-CHSH parameter $\mathcal{B}$ is provided in Appendix A. The behavior of the second-order correlation functions \( g^{2}_{a_1}(0) \) and \( g^{2}_{a_2}(0) \) across different values of normalized detuning \( \Delta/\kappa \), Kerr nonlinearity \( U/\kappa \), and intercavity coupling strength \( J \) are plotted in Fig. \ref{fig_4} using the analytical expressions of Eq. (\ref{eq22}) and Eq. (\ref{eq23}). Additionally, we demonstrate that the nonclassicality witness \( \mathcal{C} \),  greater than unity ($\log\mathcal{C}>0$), is perfectly captured by the yellow curve using Eq. (\ref{eq26}), representing the CSI violation that occurs with the single-photon blockade. These results exactly match the previously obtained numerical analysis from the master equation approach. 

Fig. \ref{fig_4}(a) shows a moderate antibunching with a slight dip in \( g^{2}_{a_1}(0) \) around resonance (\( \Delta = -0.04\kappa \)) for the weak Kerr coefficient $U = 0.09\kappa$. It can be observed that the antibunching effect is different for the two modes $a_1$ and $a_2$ due to their unequal driving settings $E_1 = 0.1\kappa, E_2 = 0.01\kappa$. These characteristics can be easily tuned by setting different driving conditions $E_1/E_2$. The nonclassicality witness \( \mathcal{C} \) exhibits small amplitude oscillations at the exact detuning conditions, indicating quantum interference effects between the cavities. The corresponding density plot in Fig.\ref{fig_4}(b) reveals two symmetric dark spots near the weak nonlinearity, \( U/\kappa \ll 1 \) where \( g^{2}_{a_1}(0) \) is way below the classical limit. These features imply that even with considerable weak nonlinearity, the detuning near resonance \( \Delta \) can facilitate nonclassical correlations, driven predominantly by intercavity coupling, taken higher than cavity linewidth \( J > \kappa \). This is widely known as the unconventional photon blockade mechanism.

As the Kerr coefficient increases to $U=0.5\kappa$, the resonance dip of \( g^{(2)}_{a_1}(0) \) shifts to $\Delta/\kappa= 0.40\kappa$, indicating enhanced antibunching due to strong photon-photon interaction owing to the Kerr effect, as shown in Fig.\ref{fig_4}(c). The solid blue curve of \( g^{2}_{a_2}(0) \) follows similar features with noticeable separation, reflecting mode asymmetry induced by the unequal driving applied. Here, the non-classicality witness shows a prominent oscillatory behavior near the resonance, confirming the onset of stronger quantum correlations. These observations are further validated in the contour plot of Fig.\ref{fig_4}(d), where more pronounced and sharper antibunching patterns appear around \( \Delta \approx \pm U \) with the coupling $J=\kappa$. Fig.\ref{fig_4}(e) highlights the most strongly nonlinear regime, where the dip reaches $\ll1$, indicating ideal photon blockade. Also, Fig.\ref{fig_4}(f) reveals the pronounced diagonal trench for high nonlinearity but weak coupling $J<\kappa$. This trench follows the condition \( \Delta \approx U \), where the single-photon resonance is favored while higher-order excitations are suppressed due to the anharmonicity from nonlinearity. Notably, the sub-Poissonian statistics are visible along the diagonal of Fig. \ref{fig_4}(f), an interesting feature also observed previously in photon filtering \cite{sanchez2014violation, darsheshdar2021photon}. This behavior suggests that the energy of the sub-Poissonian fields does not necessarily depend upon specific Fock state energies; rather, it is sufficient for their total energy to satisfy the resonance condition $\Delta-U=0$. This feature is due to the two-photon joint excitation through the virtual state $\ket{11}$, rather than the direct transition to $\ket{20}$ or $\ket{02}$.  The collective excitation through the virtual state results in destructive interference, yielding the single photon pair associated with the Fock state $\ket{11}$.

\begin{figure}[htbp]
\includegraphics[width=\linewidth]{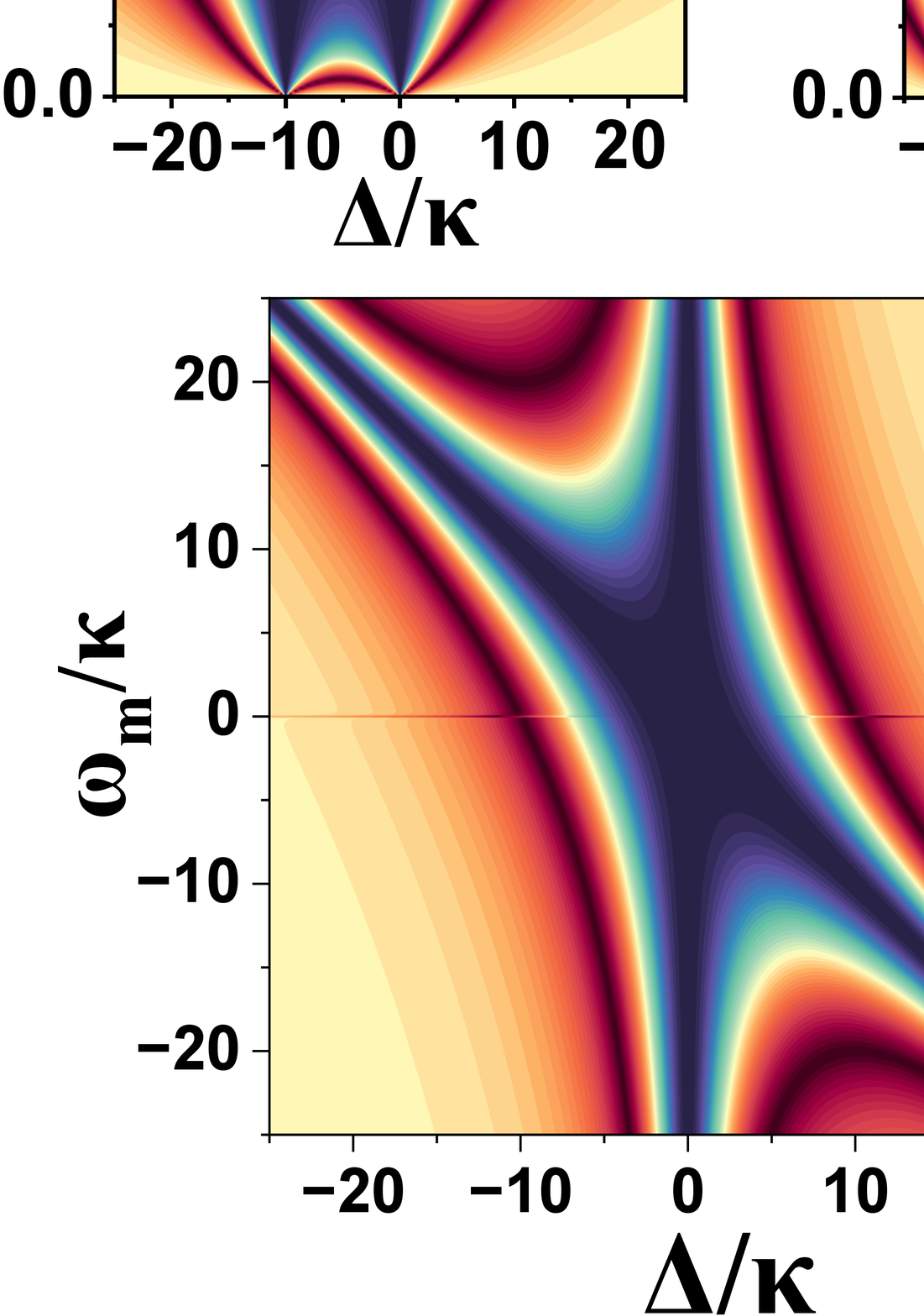}
\caption{Contour plots of the non-classicality witness $\mathcal{C}$ (left of each block) and Bell-CHSH parameter $\mathcal{B}$ (right of each block) as functions of normalized detuning $\Delta/\kappa$ and Kerr nonlinearity $U/\kappa$, for different regimes of inter-cavity coupling (a) $J = 0.5\kappa$, (b) $J = \kappa$, (c) $J = 1.5\kappa$ and (d) $J = 2\kappa$. The cyan contour lines in the plots represent the classicality threshold $\mathcal{C} = 1$ and $\mathcal{B} = 2$.}
\label{fig_5}
\end{figure}

To complement the discussion of the second-order correlation, we analyze the non-classicality and non-locality through the violation of the CSI and the Bell-CHSH inequality. The left figures of each block of Fig. \ref{fig_5} displays density plots of the CSI witness \( \mathcal{C} \) as functions of detuning \( \Delta/\kappa \) and Kerr nonlinearity \( U/\kappa \) for increasing values of intercavity coupling strengths i.e. (a) \( J = 0.5\kappa \), (b) \( J = \kappa \), (c) \( J = 1.5\kappa \), and (d) \( J = 2\kappa \). The CSI violations are indicative of nonclassical correlations between the two sub-Poissonian fields. From weak-to-strong coupling regime in Fig. \ref{fig_5}(a)-(d), the CSI violation is changing from marginal to confined to narrow bands. As \( J \) increases, broader and stronger CSI-violating regions emerge. The onset of violation closely follows the diagonal line \( \Delta \approx U \), consistent with the resonance condition for single-photon blockade observed in Fig. \ref{fig_4}. The optimum tuning of \( J \) reveals how coupling strength acts as a critical control parameter for enabling and enhancing quantum correlations. While Kerr nonlinearity \( U \) is essential for generating photon blockade and anharmonicity, it is the intercavity tunneling that facilitates delocalization and interference between antibunching optical modes, leading to entanglement and nonlocal effects.

It is anticipated that the characteristics of CHSH violation will be equivalent to those of CSI. To verify this, we plot the Bell-CHSH violation factor $\mathcal{B}$ of Eq. (\ref{eq27}) in the right figures of each block of Fig. \ref{fig_5}, where $\mathcal{B}>2$ is observed only in specific regions of the opposite quadrant. The Bell-CHSH violation captures the true nonclassical correlations upon tuning the coupling constant, showing the spatially extended nature of CSI. The maximum observed violation in Fig. \ref{fig_5}(a) approaches $\mathcal{B}\approx2.7$, which is close to the upper limit known as Tsirelson's bound. Moreover, clear boundaries of $\mathcal{B}\approx2$ are represented by the cyan lines, eliminating the incoherency encountered in CSI. It is interesting to note that, for coupling $J=2\kappa$, the nonlocal characteristics completely disappear, with the CHSH parameter $\mathcal{B}\lesssim2$. But in case \( J \lesssim \kappa \), the Bell violation appears to be stronger in small isolated regions diagonally. The violation becomes more prominent and clearly correlates with regions of strong CSI violation and sub-Poissonian photon statistics. This observation supports the trend identified in the earlier results where increasing \( U \) sharpens the photon blockade, while increasing \( J \) enhances the spatial coherence and mode entanglement necessary for Bell inequality violation. Thus, a cooperative balance between strong nonlinearity and coherent tunneling is essential for realizing strongly nonclassical and nonlocal states.

\section{Fidelity of Bell-type states}\label{sec6}
\begin{figure}
    \centering
    \includegraphics[width=\linewidth]{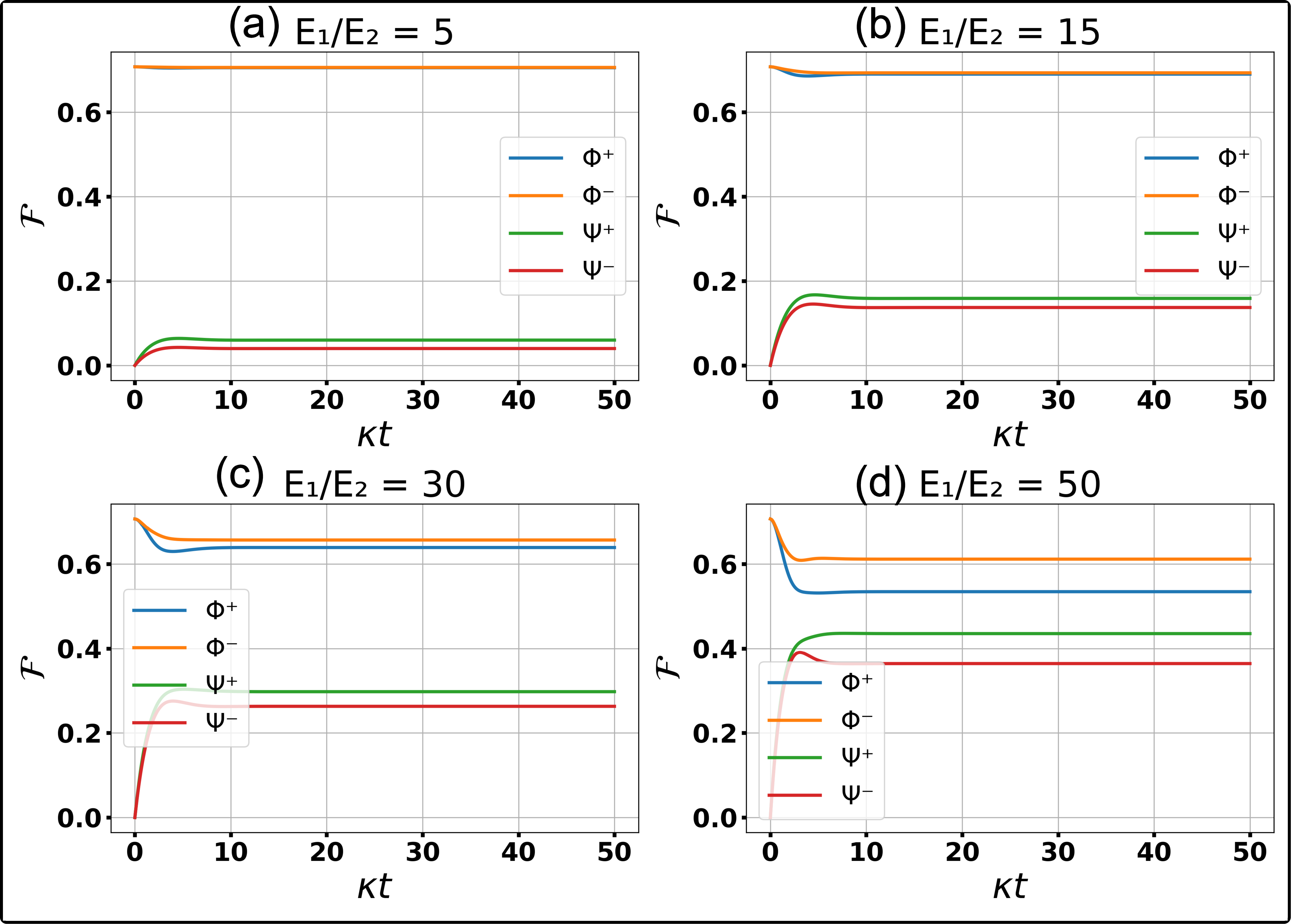}
    \caption{
Time evolution of the fidelity \( \mathcal{F} \) with respect to the four Bell states \( \ket{\Phi^+} \), \( \ket{\Phi^-} \), \( \ket{\Psi^+} \), and \( \ket{\Psi^-} \), for varying drive ratios \( E_1/E_2 \) with fixed inter-cavity coupling \( J = \kappa/2 \) and detuning condition \(\Delta=U=\kappa\). 
Panels (a)--(d) correspond to \( E_1/E_2 = 5,\,15,\,30,\,50 \), respectively. The state \( \ket{\Phi^\pm} \) consistently exhibits the highest fidelity, while increased asymmetry in the drives enhances the fidelity of \( \ket{\Psi^\pm} \). 
}
    \label{fig_6}
\end{figure}

To further characterize the quantum correlations between the two coupled optical modes, we analyze the projection of the system’s steady state onto the four maximally entangled Bell states, $
\ket{\Phi^\pm} = \frac{1}{\sqrt{2}} \left( \ket{00} \pm \ket{11} \right)$ and
$\ket{\Psi^\pm} = \frac{1}{\sqrt{2}} \left( \ket{01} \pm \ket{10} \right)$ where the logical basis \( \ket{n_in_j} \) corresponds to photon occupation in the respective cavity modes \( a_1 \) and \( a_2 \).

Fig. \ref{fig_6} shows the time evolution of the fidelity \( \mathcal{F}(t) = \mathrm{Tr} \left[ \rho(t) \ket{\psi_{\text{Bell}}}\bra{\psi_{\text{Bell}}} \right] \) with respect to these Bell states, obtained by solving the Eq. \ref{eq4} starting from the vacuum state $\ket{00}$. The parameters of our interest are the drive asymmetry corresponding to increasing ratios \( E_1/E_2 = 5,\,15,\,30,\,50 \), with the inter-cavity coupling fixed at \( J = \kappa \) and detuning condition $\Delta=U$. As the ratio \( E_1/E_2 \) increases, the fidelity of the Bell state \( \ket{\Phi^-} \) remains dominant in all cases, suggesting that the system naturally evolves toward this state under asymmetric driving. Notably, \( \ket{\Phi^+} \) also loses fidelity for higher imbalance, indicating that the drive asymmetry can be tuned to selectively enhance certain Bell-like components in the steady state, not altogether. But unlike deterministic entanglement protocols (e.g., Rabi oscillation or conditional logic gates) \cite{kamimaki2023deterministic}, the underlying mechanism here is inherently probabilistic. Our model explores photon antibunching in coupled optomechanical systems, where photon blockade arises from multi-path quantum interference, a phenomenon rooted in the interplay of Kerr nonlinearity and inter-cavity coupling. While both CSI and Bell-CHSH inequality violations have been observed in this parameter regime, they do not directly imply high-fidelity projection onto maximally entangled Bell states. Rather, these violations signal the presence of quantum correlations, which may demonstrate entanglement but also extend to broader nonclassical correlations between the modes. The CHSH violation specifically captures nonlocal correlations but does not guarantee that the steady state lies close to a maximally entangled Bell state in Hilbert space. The fidelity values $\mathcal{F}\leqslant0.7$ suggest that although quantum correlations are present, the system does not deterministically prepare a Bell state. This opens the possibility for heralded entanglement schemes \cite{wang2021accelerated, dhara2023entangling}, where a measurement on one subsystem (e.g., photon detection in one mode) projects the other subsystem into a desired entangled state. Such heralding could potentially increase the fidelity of Bell state generation in this non-deterministic, interference-driven regime. However, this analysis demonstrates that drive asymmetry provides a powerful tool to control the entangled output states in coupled nonlinear systems, while also highlighting the distinction between quantum correlations (as captured by inequality violations) and high-fidelity entanglement generation.

\section{Conclusion}

In this work, we have demonstrated photon antibunching in a weakly perturbed coupled optomechanical cavity, demonstrating the simultaneous violation of both the Cauchy-Schwarz and Bell-CHSH inequalities. These nonclassical photon statistical effects become prominent when the system operates near resonance between optical cavity detunings, in the presence of a tunable coupling strength and Kerr nonlinearity. The generation of such strongly nonclassical states requires an optomechanical coupling strength $g \ll \omega_m$, which is attainable with current experimental techniques \cite{rabl2011photon, nunnenkamp2011single}. Although achieving single-photon strong coupling ($g \geqslant \kappa$) remains more challenging, feasible approaches have been demonstrated, such as in \cite{groblacher2009observation}. Alternative routes to enhancing photon-phonon interaction include the use of quartz crystals \cite{tomasella2025strong}. However, platforms like photonic crystals \cite{chan2011laser} and electromechanical devices \cite{teufel2011circuit} still fall short of reaching the required coupling regimes.
Additionally, mechanical elements such as SiN nanoscale trampolines \cite{wei2019controllable} or photonic crystal nanobeam resonators \cite{riedinger2016non} offer promising architectures. These structures can support high-quality mechanical breathing modes with $\omega_m / \gamma > 10^6$, while maintaining weak Kerr nonlinearities and low thermal occupancies at cryogenic temperatures—both essential for observing classical inequality violations.

Overall, the theoretical framework relies on number-squeezing effects in coherent optical fields under weak driving conditions. The results presented here pave the way for observing macroscopic quantum phenomena using on-chip light–matter interfaces. Furthermore, this approach can be generalized to a broad class of hybrid systems, providing a feasible platform for testing quantum behavior with existing technologies.
\\

\backmatter
\bmhead{Acknowledgements}

The authors acknowledge the technical discussions and inputs provided by  Prof. A.K. Sarma and Dr. R. Nongthombam of IIT Guwahati. Authors also acknowledge funding support from the National Quantum Mission, an initiative of the Department of Science and Technology, Government of India.

\bmhead{Data availability}
Data available from the corresponding author upon a reasonable request.

\begin{appendices}

\section{Analytical expression of Bell-CHSH inequality}\label{secA1}

In the appendix, we compute the Bell-CHSH test parameter $\mathcal{B}$ in the coupled optomechanical system, which positions the violation of the Cauchy-Schwarz (CS) inequality as a preliminary indicator for a more stringent test of nonclassicality. To this end, we adopt a protocol analogous to those used in quantum optical Bell-type experiments, where intensity measurements of the two optical modes $\hat{a}_1$ and $\hat{a}_2$ are performed after interference with phase-controlled local oscillators (LOs). These mixed fields are directed to four photodetectors, yielding binary outcomes per measurement event, and allow for the construction of the correlation functions necessary to evaluate the Bell-CHSH parameter.

The experimental scheme is conceptually aligned with proposals and implementations such as those in \cite{vivoli2016proposal, marinkovic2018optomechanical}, where optomechanical modes are measured through homodyne or heterodyne detection involving tunable LOs. The generated fields, arising from optomechanically induced parametric interactions, are analyzed through their detected intensities—denoted by $\langle I_{A_+} \rangle$, $\langle I_{A_-} \rangle$, $\langle I_{B_+} \rangle$, and $\langle I_{B_-} \rangle$—which depend on adjustable LO polarization angles $\Theta$ and $\Phi$. These intensity outcomes allow for the evaluation of expectation values of correlation functions between measurement settings on subsystems $A$ and $B$. To realize the necessary entangled-like correlations from the optical modes, a 50:50 beam-splitter transformation is applied to the original modes $\hat{a}_1$ and $\hat{a}_2$, yielding output modes suitable for Bell-type measurements. This transformation preserves coherence and allows for tunable mode mixing, critical for revealing quantum correlations beyond classical limits, given as

\begin{eqnarray}
\begin{pmatrix}
    A_{+}
    \\
    A_{-}\end{pmatrix} = \begin{pmatrix}
        \cos{\Theta} && \sin{\Theta}\\
        -\sin{\Theta} && \cos{\Theta}
    \end{pmatrix} \begin{pmatrix}
        \hat{a}_1
        \\
         \hat{a}_2
    \end{pmatrix},
    \label{A1}\\
    \begin{pmatrix}
    B_{+}
    \\
    B_{-}\end{pmatrix} = \begin{pmatrix}
        \cos{\Phi} && -\sin{\Phi}\\
        \sin{\Phi} && \cos{\Phi}
    \end{pmatrix}  \begin{pmatrix}
        \hat{a}_1
        \\
         \hat{a}_2
    \end{pmatrix}.
    \label{A2}
\end{eqnarray}
Finally, the four correlated pairs of the field intensity measurements i.e. $\langle I_{A_+}I_{B_+}\rangle$, $\langle I_{A_+}I_{B_-}\rangle$, $\langle I_{A_-}I_{B_+}\rangle$ and $\langle I_{A_-}I_{B_-}\rangle$ are done using a Mach-Zehnder
interferometer and the following Bell-CHSH test parameter is estimated
    \begin{eqnarray}
    E(\Theta,\Phi)&=&\frac{\langle I_{A_+}I_{B_+}\rangle+\langle I_{A_-}I_{B_-}\rangle-\langle I_{A_+}I_{B_-}\rangle-\langle I_{A_-}I_{B_+}\rangle}{\langle I_{A_+}I_{B_+}\rangle+\langle I_{A_-}I_{B_-}\rangle+\langle I_{A_+}I_{B_-}\rangle+\langle I_{A_-}I_{B_+}\rangle},
    \nonumber\\
     &=& \frac{\langle(I_{A_+}-I_{A_-})(I_{B_+}-I_{B_-})\rangle}{\langle(I_{A_+}+I_{A_-})(I_{B_+}+I_{B_-})\rangle},
     \nonumber\\ &=& \frac{\langle:(A^\dagger_{+}A_{+}-A^\dagger_{-}A_{-})(B^\dagger_{+}B_{+}-B^\dagger_{-}B_{-}):\rangle}{\langle:(A^\dagger_{+}A_{+}+A^\dagger_{-}A_{-})(B^\dagger_{+}B_{+}+B^\dagger_{-}B_{-}):\rangle},
     \label{A3}
     \end{eqnarray}
    where $\langle::\rangle$ denotes the normal order of the operators. The correlations of the detected modes are found as
    \begin{eqnarray}
A^\dagger_{+}A_{+}+A^\dagger_{-}A_{-} &=&\begin{pmatrix}
    \hat{a}_1^\dagger & \hat{a}_2^\dagger
    \end{pmatrix}
    \begin{pmatrix}
        \cos{\Theta} & \sin{\Theta}\\
        -\sin{\Theta} &  \cos{\Theta}
    \end{pmatrix}\times
    \begin{pmatrix}
        \cos{\Theta} & \sin{\Theta}\\
        -\sin{\Theta} &  \cos{\Theta}
    \end{pmatrix}^T
    \begin{pmatrix}
    \hat{a}_1^\dagger
    \\ \hat{a}_2^\dagger
    \end{pmatrix} \nonumber\\&&
    = \hat{a}_1^\dagger \hat{a}_1+ \hat{a}_2^\dagger \hat{a}_2,
    \label{A4}
    \end{eqnarray}
    \begin{eqnarray}
    B^\dagger_{+}B_{+}+B^\dagger_{-}B_{-} &=&\begin{pmatrix}
    \hat{a}_1^\dagger & \hat{a}_2^\dagger
    \end{pmatrix}
    \begin{pmatrix}
        \cos{\Phi} & -\sin{\Phi}\\
        \sin{\Phi} &  \cos{\Phi}
    \end{pmatrix}\times
    \begin{pmatrix}
        \cos{\Phi} & -\sin{\Phi}\\
        \sin{\Phi} &  \cos{\Phi}
    \end{pmatrix}^T
    \begin{pmatrix}
    \hat{a}_1^\dagger
    \\ \hat{a}_2^\dagger
    \end{pmatrix} \nonumber\\&&
    = \hat{a}_1^\dagger \hat{a}_1+ \hat{a}_2^\dagger \hat{a}_2,
    \label{A5}
    \end{eqnarray}
    \begin{eqnarray}
    A^\dagger_{+}A_{+}-A^\dagger_{-}A_{-} &=&\begin{pmatrix}
    \hat{a}_1^\dagger & \hat{a}_2^\dagger
    \end{pmatrix}
    \begin{pmatrix}
        \cos{\Theta} & \sin{\Theta}\\
        \sin{\Theta} &  -\cos{\Theta}
    \end{pmatrix}\times
    \begin{pmatrix}
        \cos{\Theta} & \sin{\Theta}\\
        -\sin{\Theta} &  \cos{\Theta}
    \end{pmatrix}
    \begin{pmatrix}
    \hat{a}_1^\dagger
    \\ \hat{a}_2^\dagger
    \end{pmatrix},
    \nonumber\\
    &=& (\hat{a}_1^\dagger \hat{a}_1- \hat{a}_2^\dagger\hat{a}_2)\cos{2\Theta}+(\hat{a}_1^\dagger \hat{a}_2+ \hat{a}_2^\dagger\hat{a}_1)\sin{2\Theta},
    \label{A7}
    \end{eqnarray}
    \begin{eqnarray}
    B^\dagger_{+}B_{+}-B^\dagger_{-}B_{-} &=&\begin{pmatrix}
    \hat{a}^\dagger & \hat{b}^\dagger
    \end{pmatrix}
    \begin{pmatrix}
        \cos{\Phi} & -\sin{\Phi}\\
        -\sin{\Phi} &  -\cos{\Phi}
    \end{pmatrix}\times
    \begin{pmatrix}
        \cos{\Phi} & -\sin{\Phi}\\
        \sin{\Phi} &  \cos{\Phi}
    \end{pmatrix}
    \begin{pmatrix}
    \hat{a}^\dagger
    \\ \hat{b}^\dagger
    \end{pmatrix},
    \nonumber\\
    &=& (\hat{a}_1^\dagger \hat{a}_1- \hat{a}_2^\dagger\hat{a}_2)\cos{2\Phi}-(\hat{a}_1^\dagger \hat{a}_2+ \hat{a}_2^\dagger\hat{a}_1)\sin{2\Phi}.
    \label{A8}
    \end{eqnarray}
    The denominator and numerator of Eq. (\ref{A3}) is simplified as
    \begin{eqnarray}
        \langle:(A^\dagger_{+}A_{+}+A^\dagger_{-}A_{-})(B^\dagger_{+}B_{+}+B^\dagger_{-}B_{-}):\rangle
        =\langle \hat{a}_1^{\dagger2}\hat{a}_1^2\rangle +\langle \hat{a}_2^{\dagger2}\hat{a}_2^2\rangle
        +2\langle \hat{a}_1^{\dagger}\hat{a}_2^{\dagger}\hat{a}_2\hat{a}_1\rangle,
        \label{A6}
    \end{eqnarray}
and
\begin{eqnarray}
    \langle:(A^\dagger_{+}A_{+}-A^\dagger_{-}A_{-})(B^\dagger_{+}B_{+}-B^\dagger_{-}B_{-}):\rangle
  &=&\Bigl\langle:(\hat{a}_1^\dagger \hat{a}_1- \hat{a}_2^\dagger\hat{a}_2)\cos{2\Theta}
   +(\hat{a}_1^\dagger \hat{a}_2+ \hat{a}_2^\dagger\hat{a}_1)\sin{2\Theta}
   \nonumber\\&&\times(\hat{a}_1^\dagger \hat{a}_1- \hat{a}_2^\dagger\hat{a}_2)\cos{2\Phi}
    -(\hat{a}_1^\dagger \hat{a}_2+ \hat{a}_2^\dagger\hat{a}_1)\nonumber\\&&\sin{2\Phi}:\Bigr\rangle.
    \label{A9}
\end{eqnarray}
The final expression of Eq.(\ref{A3}) becomes
        \begin{eqnarray}
    E(\Theta,\Phi)&=&\frac{1}{\langle \hat{a}_1^{\dagger2}\hat{a}_1^2\rangle +\langle \hat{a}_2^{\dagger2}\hat{a}_2^2\rangle
        +2\langle \hat{a}_1^{\dagger}\hat{a}_2^{\dagger}\hat{a}_2\hat{a}_1\rangle}\times\biggr[\langle(\hat{a}_1^\dagger \hat{a}_1- \hat{a}_2^\dagger\hat{a}_2)^2\rangle\cos{2\Theta}\cos{2\Phi}\nonumber\\&&-\langle(\hat{a}_1^\dagger \hat{a}_2+ \hat{a}_2^\dagger\hat{a}_1)^2\rangle
    \times\sin{2\Theta}\sin{2\Phi}
    -\langle\hat{a}_1^\dagger \hat{a}_1- \hat{a}_2^\dagger\hat{a}_2\rangle\langle\hat{a}_1^\dagger \hat{a}_2+ \hat{a}_2^\dagger\hat{a}_1\rangle\cos{2\Theta}\sin{2\Phi}
    \nonumber\\&&+\langle\hat{a}_1^\dagger \hat{a}_2+ \hat{a}_2^\dagger\hat{a}_1\rangle\langle\hat{a}_1^\dagger\hat{a}_1-\hat{a}_2^\dagger\hat{a}_2\rangle\sin{2\Theta}\cos{2\Phi}\biggr].
    \nonumber\\&&
    \label{A10}
\end{eqnarray}

The Bell-CHSH parameter can be tested by violating the following inequality
\begin{eqnarray}
\mathcal{B}=\abs{E(\Theta,\Phi)+E(\Theta',\Phi')+E(\Theta',\Phi)
-E(\Theta,\Phi')}\leqslant2,
\nonumber\\
\label{A11}
\end{eqnarray}
By systematically varying the LO phases, the Bell-CHSH correlation functions are computed with the maximum violation obtained by the standard choice of polarization angles chosen as $\Theta=0,\Phi=\pi/8,\Theta'=\pi/4,\Phi'=3\pi/8$. In terms of the original OM modes Eq. (\ref{A11}) can be rewritten as
\begin{eqnarray}
    \mathcal{B}=\sqrt{2}\abs{\frac{\langle \hat{a}_1^{\dagger2}\hat{a}_1^2\rangle+\langle \hat{a}_2^{\dagger2}\hat{a}_2^2\rangle-\langle \hat{a}_1^{\dagger2}\hat{a}_2^2\rangle-\langle \hat{a}_2^{\dagger2}\hat{a}_1^2\rangle
    -4\langle \hat{a}_1^{\dagger}\hat{a}_2^{\dagger}\hat{a}_2\hat{a}_1\rangle}{\langle \hat{a}_1^{\dagger2}\hat{a}_1^2\rangle+\langle \hat{a}_2^{\dagger2}\hat{a}_2^2\rangle
    +2\langle \hat{a}_1^{\dagger}\hat{a}_2^{\dagger}\hat{a}_2\hat{a}_1\rangle}}.
    \nonumber\\
    \label{A12}
\end{eqnarray}
It is now straightforward to calculate the Bell-CHSH violation parameter $\mathcal{B}$ by replacing the expectation values of the operators $\hat{a}_1$ and $\hat{a}_2$ with the transition probabilities obtained from Eq. (\ref{eq26}) to Eq. (\ref{eq27}).\\

\end{appendices}

\bibliography{sn-bibliography}

\end{document}